\documentclass[lettersize,journal]{IEEEtran}
\usepackage{amsmath,amsfonts}
\usepackage{algorithmic}
\usepackage{algorithm}
\usepackage{array}
\usepackage[caption=false,font=normalsize,labelfont=sf,textfont=sf]{subfig}
\usepackage{textcomp}
\usepackage{stfloats}
\usepackage{url}
\usepackage{verbatim}
\usepackage{graphicx}
\usepackage{cite}
\usepackage{makecell} 
\usepackage{multirow} 
\usepackage[T1]{fontenc} 

\begin{document}

\title{Dynamic Prediction of Full-Ocean Depth SSP by Hierarchical LSTM: An Experimental Result}

\author{Jiajun Lu, Wei Huang~\IEEEmembership{Member,~IEEE,} Hao Zhang~\IEEEmembership{Senoir Member,~IEEE}
\thanks{Wei Huang, Jiajun Lu and Hao Zhang are with the Faculty of Information Science and Engineering, Ocean University of China, Qingdao, China (e--mail: hw@ouc.edu.cn, zhanghao@ouc.edu.cn) (Corresponding author: Wei Huang and Hao Zhang.)}
\thanks{Manuscript received October XX, 2023; revised XX XX, 2023.}}

\markboth{IEEE Geoscience and Remote Sensing Letters,~Vol.~XX, No.~X, October~2023}%
{Shell \MakeLowercase{\textit{et al.}}: A Sample Article Using IEEEtran.cls for IEEE Journals}


\maketitle

\begin{abstract}
SSP distribution is an important parameter for underwater positioning, navigation and timing (PNT) because it affects the propagation mode of underwater acoustic signals. To accurate predict future sound speed distribution, we propose a hierarchical long short--term memory (H--LSTM) neural network for future sound speed prediction, which explore the distribution pattern of sound velocity in the time dimension. To verify the feasibility and effectiveness, we conducted both simulations and real experiments. The ocean experiment was held in the South China Sea in April, 2023. Results show that the accuracy of the proposed method outperforms the state--of--the--art methods.
\end{abstract}


\section{Introduction}   
\IEEEPARstart{U}{nderwater} sound speed distribution is one of the most important parameters for underwater positioning, navigation and timing (PNT) because it affects the propagation mode of underwater acoustic signals \cite{Wang2013Marine}. With the increasing demand for precision performance in PNT, it is necessary to quickly and accurately obtain the regional sound speed distribution, and even predict the future sound speed distribution, so as to predict the future sound field distribution.

The acquisition of ocean sound speed profile (SSP) mainly includes SSP measurement method and SSP inversion method. The SSP can be measured directly by using the sound velocity profiler (SVP) \cite{Xu2012SVP}, or can be measured indirectly by conductivity, temperature and depth profiler (CTD) according to \cite{Yuan2009CTD,wang2014CTD,CTD2020} and expendable CTD profiler (XCTD) \cite{XCTD2023} combined with empirical sound speed formula. However, the SSP measurement method usually takes a long time \cite{Huang2023Fast}. 

To fast obtain sound speed distribution, ocean SSPs inversion methods have been widely studied. The traditional methods for SSP inversion are mainly divided into three categories: matched field processing (MFP), compressed sensing (CS), and deep learning (DL) methods. Most of the SSP inversion methods focus on spatial SSP construction, which mainly relies on real-time sonar observation data. There are relatively few studies on SSP prediction.

In 1979, Munk and Wunsch first proposed the concept of SSP in \cite{munk1979ocean} and \cite{munk1983ocean-2}, and put forward the idea of inverting the SSP through signal propagation time. In 1995, Tolstoy proposed a MFP for SSP inversion \cite{tolstoy1991acoustic}, where an effective solution is provided for the inability to establish a mapping relationship from sound field distribution to SSP. In 2000, Shen \textit{et al.} demonstrated the feasibility of inversion of SSPs by using empirical orthogonal functions (EOFs) in shallow sea area \cite{Shen1999EOF}. In 2006, Jain \textit{et al.} proposed a method for estimating SSPs based on artificial neural networks in \cite{Jain2006ANN}. However, these methods does not gain good SSP inversion accuracy.

To improve the accuracy of SSP inversion, Han \textit{et al.} proposed an improved SSPs estimation method based on the traditional EOF \cite{Han2009EOF}. Bianco et~al. proposed a CS-based framework for SSP inversion \cite{Bianco2016CS}. Our previous work proposed a deep-learning model for SSP inversion \cite{Huang2019Collaborating} that also improves the real-time performance. The above methods have effectively improved the inversion accuracy of SSPs, but they rely on real--time ocean observation data during the inversion process. Recently, Li \textit{et al.} proposed a self-organizing map (SOM) neural network that combines surface sound speed for SSP estimation in \cite{Li2022Inversion}, which achieves the construction of SSP without sonar observation data, but it is unable to predict future sound speed distribution.

The aforementioned inversion methods can achieve relatively accurate spatial dimension SSPs inversion, but they lack the ability to capture the law of sound speed distribution changing with time. To tackle the problem of SSP prediction in the time dimension, we propose a future sound speed dynamic prediction method based on hierarchical long short--term memory (H--LSTM) neural networks. The model can dynamically adjust the time step size of the prediction based on the different characteristics of training datasets, which determines the time resolution. Taking the historical SSP data in the spatial area where the prediction tasks are located as a reference, we first process the historical sound speed distribution data in layers and set up different H-LSTM neural network models for different depth layers, then train and predict the sound speed value in different depth layers, finally combine the prediction results of each layer of H-LSTM model to form full--ocean depth SSP.

\section{Methodology}    
\subsection{Structure of H--LSTM}
Long short-term memory (LSTM) neural network is a special type of recurrent neural network (RNN) that can solve the problems of gradient vanishing and explosion in RNN, and performs well in time series prediction models \cite{Hochreiter1997LSTM}. In this letter, we propose a hierarchical long short-term memory (H-LSTM) neural network for accurate SSP prediction. The core idea is the hierarchical processing of data. Due to the different characteristics of data at different layers, the data are layered by depth, and corresponding H-LSTM models are defined for different depth layers. Then, the time series data in each depth layer are trained and predicted.

\begin{figure*}[!t]
	\centering
	\includegraphics[width=5in]{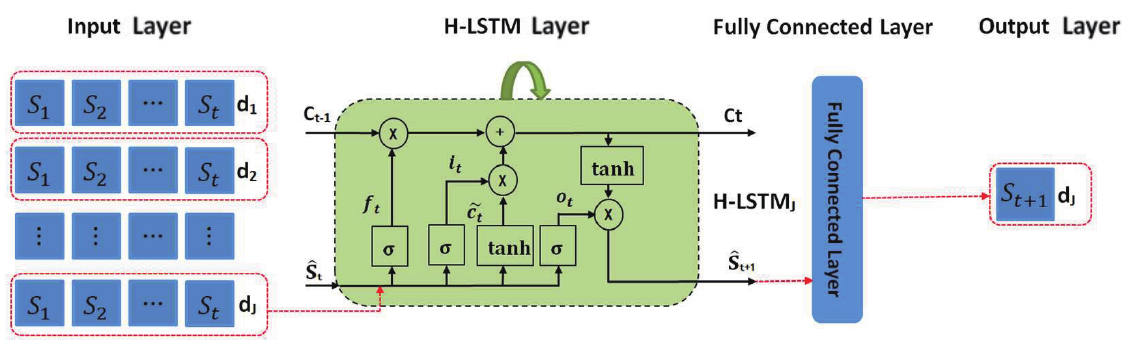}
	\caption{The structure of H-LSTM.}
	\label{fig_1}
\end{figure*}

The structure of H-LSTM is shown in Fig.~\ref{fig_1}, which mainly contains an input layer, an H-LSTM layer, a fully connected layer, and an output layer. Before training, the data is first layered based on depth and sorted in time dimension. Then, they are normalized into a hierarchical sound speed distribution time standardized dataset. For a given time series sound speed data, the hierarchical SSP data $S_t$ at the current time is fed into the model, and cell state $C_t$ is obtained in the H--LSTM layer by:
\begin{equation} \label{eq1}
	C_{t} = f_{t}C_{t-1} + i_t\tilde{C}_t,
\end{equation}
where $i_t=\sigma(W_i[\hat{S}_t,S_t]+ b_i)$, and $\tilde{C}_t = tanh(W_C[\hat{S}_t,S_t]+ b_C)$. $i_t$ is the output of output gate, $\tilde{C}_t$ is the candidate cell status, $\sigma()$ denotes the sigmoid function, $\hat{S}_t$ is the previous estimated sound speed value for current time step, $W_i$, $W_C$, $b_i$, $b_C$ are the weight matrices and biases, respectively. Then the estimated $S_{t+1}$ for the next time-step will be:

\begin{equation}\label{eq2}
	\hat{S}_{t+1} = o_{t}tanh(C_t),
\end{equation}
where $o_t = \sigma(W_o[\hat{S}_t,S_t]+b_o)$ is the output gate, $tanh()$ is the Tanh activation function, $W_o$ and $b_o$ are the weight matrix and bias, respectively. Back propagation (BP) is performed to update all weights by calculating the error between predicted and actual sound speed values. The Loss function is given in the form of root mean square error (RMSE) by:

\begin{equation}\label{eq3}
	RMSE_{dj}=\sqrt{\left({S}_{p,dj}-{S}_{r,dj}\right)^2}.
\end{equation}
where ${S}_{p,dj}$ and ${S}_{r,dj}$ represent the predict and actual sound speed values, respectively. The purpose of adding a fully connected layer between the H-LSTM layer and the output layer is to improve the fitting performance of the model, and finally, the output layer will provide future SSP prediction results $\hat{S}_{t+1}$.

\subsection{Workflow of H-LSTM for SSP Prediction}
The specific implementation steps of the sound speed distribution prediction method includes data set preprocessing, H-LSTM neural network building, model training, model validation and SSP prediction. To better describe the prediction process of future full-ocean depth SSP, we systematically summarized the H-LSTM based future SSP prediction algorithm in Algorithm \ref{alg_1}.

\begin{algorithm}[H]
	\caption{H-LSTM Algorithm}\label{alg_1}
	\begin{algorithmic}
		\STATE 
		\STATE {\textsc{\bf{Input:}}} Normalize hierarchical temporal SSPs:
		\STATE \hspace{0.5cm}$\bf{\tilde{S}}=\begin{bmatrix}
			\tilde{s}_{t+1-nc,d1}&\tilde{s}_{t+2-nc,d1}&\cdots&\tilde{s}_{t,d1}\\
			\vdots&\vdots&\ddots&\vdots\\
			\tilde{s}_{t+1-nc,dj}&\tilde{s}_{t+2-nc,dj}&\cdots&\tilde{s}_{t,dj}
		\end{bmatrix}$  \\
		\STATE \hspace{1cm}Validation SSP with full-ocean depth:$\bf{S_{t+1}}$.\\
		\STATE {\textsc{\bf{Output:}}}Predicted SSP with full-ocean depth:$\bf{\check{S}_{t+1}}$.\\
		\STATE {\textsc{\bf{Step 1:}}}Constructing a dynamic H-LSTM neural network; \\
		\STATE {\textsc{\bf{Step 2:}}}Model training; \\
		\STATE {\textsc{\bf{Step 3:}}}Model validation (model output, output data reverse-normalizing, and RMSE calculation); \\
		\STATE {\textsc{\bf{Step 4:}}}Predicting future SSP with full-ocean depth. \\
	\end{algorithmic}
\end{algorithm}

\begin{table*}[!t] 
	\caption{DATA SOURCE\label{tab_1}}
	\centering 
	\begin{tabular}{|c|c|c|c|c|c|}
		\hline
		\multicolumn{6}{|c|}{\textbf{Argo Dataset}}\\  
		\hline
		{Study Area}&{Time Dimension}&{Temporal Resolution}&{Number of SSP}&{Full-depth}&{Layered Processing}\\
		\hline
		76.5$^{\circ}$E,29.5$^{\circ}$S&2017-2021(60 Months)&Month Mean&60&0-1975meters&unequal interval(58 layers)\\
		\hline
		\multicolumn{6}{|c|}{\textbf{Ocean Experiments Dataset}}\\
		\hline
		{Study Area}&{Time Dimension}&{Temporal Resolution}& {Number of SSP}&{Full-depth}&{Layered Processing}\\
		\hline
		116$^{\circ}$E,20$^{\circ}$N&March 27, 2023(24 Hours)&Approximately two hours mean&14&0-3500meters&equal interval(36 layers)\\
		\hline
	\end{tabular}
\end{table*}

\begin{figure*}[!htbp]   
	\centering
	\subfloat[]{\includegraphics[width=1.5in]{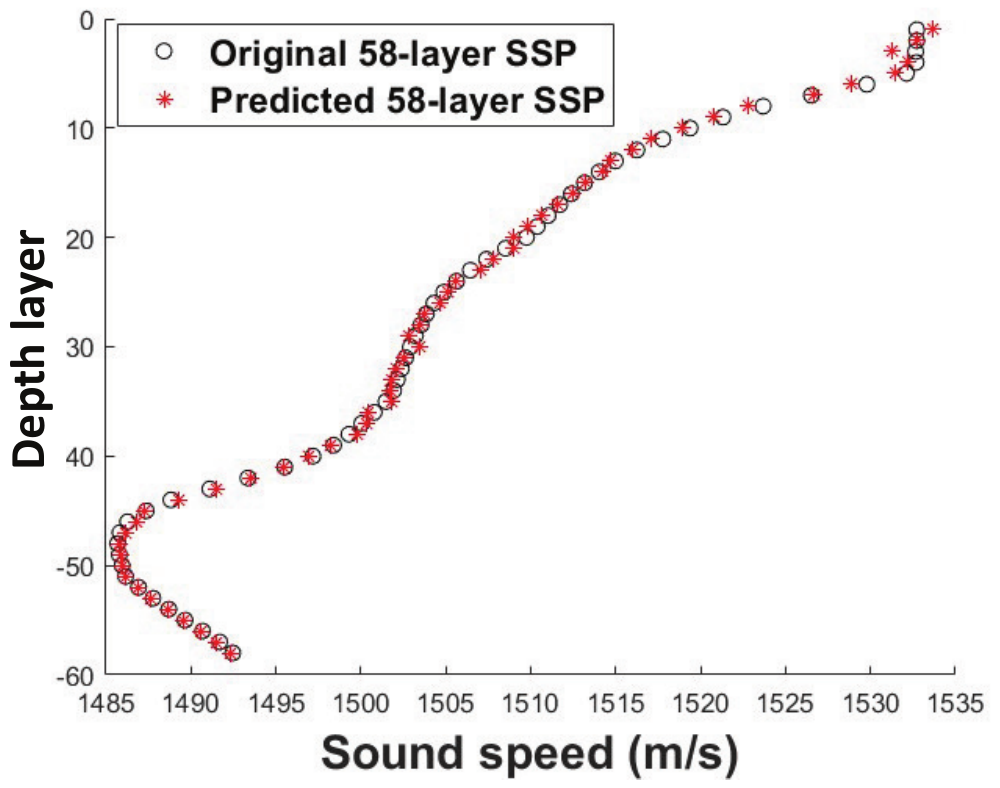}%
		\label{fig_4_1}}
	\hfil
	\subfloat[]{\includegraphics[width=1.5in]{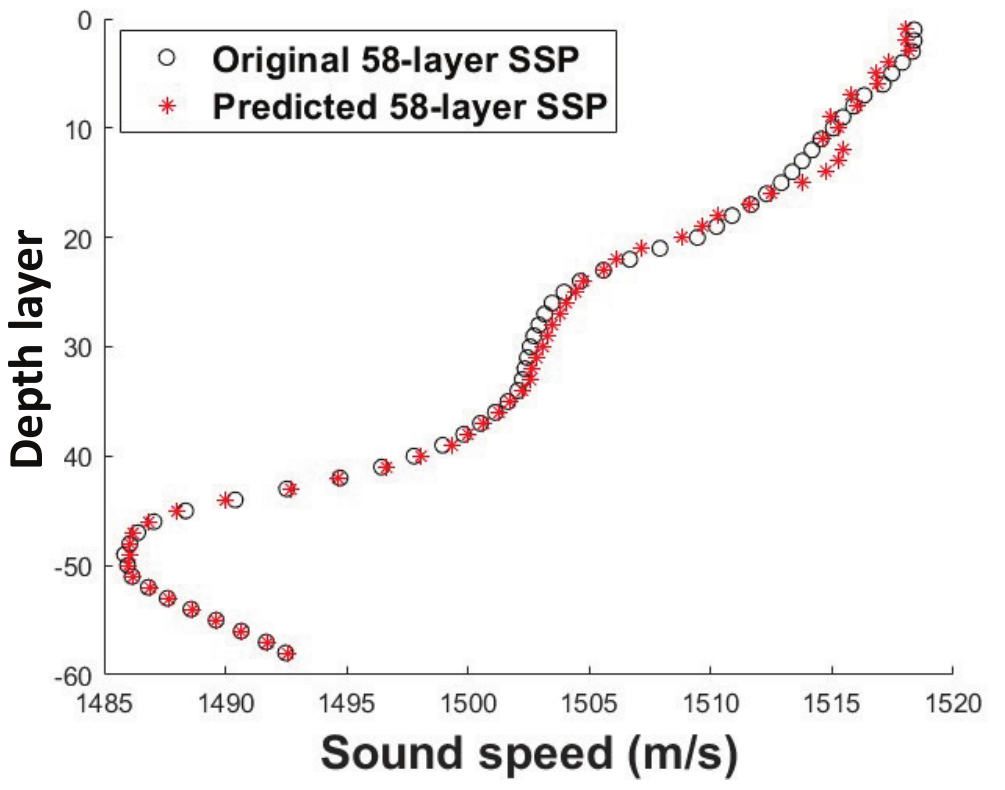}%
		\label{fig_4_2}}
	\hfil
	\subfloat[]{\includegraphics[width=1.5in]{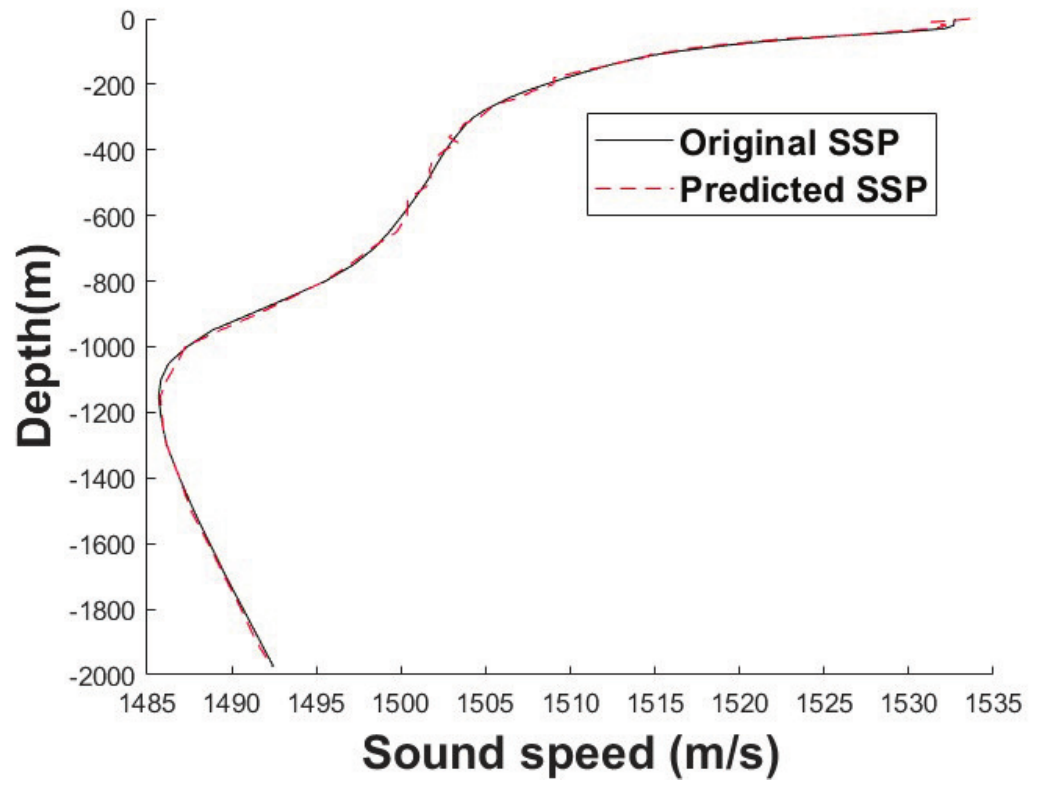}%
		\label{fig_4_3}}
	\hfil
	\subfloat[]{\includegraphics[width=1.5in]{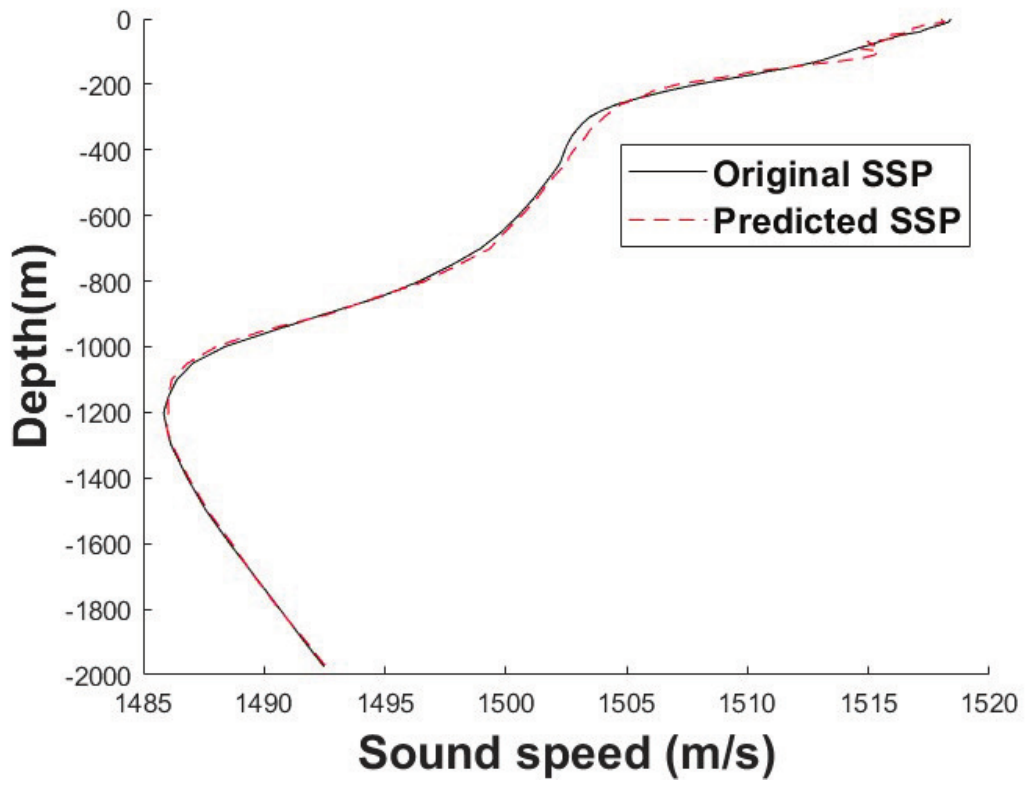}%
		\label{fig_4_4}}	
	\caption{Comparison between predicted hierarchical or full-ocean depth SSP and actual hierarchical or full-ocean depth SSP (ARGO Dataset). (a), (b) or (c), (d) respectively represent February and October 2021. (a), (b) original 58 layers' SSP. (c), (d) resampled SSP. }
	\label{fig_4}
\end{figure*}

\section{Results and Discussion}   
\subsection{Ocean Experiments}
\indent To evaluate the feasibility and effectiveness of proposed H--LSTM method for SSP prediction. We conducted deep--ocean experiments at the South Sea of China with areas of $10 km \times 10 km$ in middle April 2023, where the depth is over 3500 meters. The relevant data collection corresponding to SSP inversion lasted for a total of 3 days. 

\begin{figure}[!htbp]   
	\centering
	\subfloat[]{\includegraphics[width=1.6in]{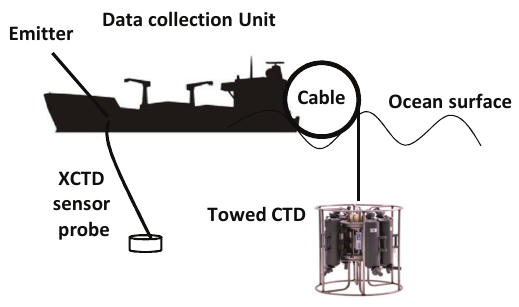}%
		\label{fig_2a}}
	\hfil
	\subfloat[]{\includegraphics[width=1.4in]{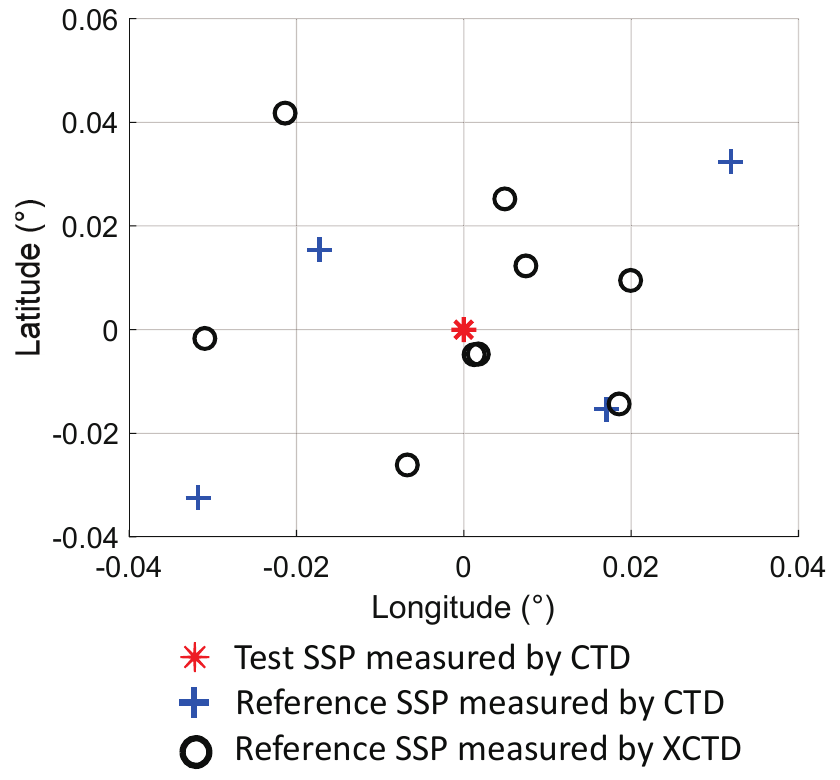}%
		\label{fig_2b}}
	\caption{Data collection. (a) Sampling by CTD and XCTD. (b) Data location.}
	\label{fig_2}
\end{figure}

\indent The system composition is shown in Fig.\ref{fig_2a}, including, a ship unit that containing a CTD, a set of expendable CTD (XCTD). SSP samples were collected by CTD and XCTD. The real--time position of the ship was located through the global positioning system (GPS), which was installed near the central axis of the ship. SSP samples were collected by CTDs and XCTDs. A full--depth of SSP was measured through ship borne CTD, the product model of which is SBE911 produced by Sea-bird Scientific \cite{SEABIRD}. Considering the high time costs of SSP measurement by CTD (almost 3 hours for once measurement with no ship movement), we used the XCTD to collect the other 13 SSPs, the model of which is HYLMT-2000 produced by \cite{HAIYAN}. XCTD provides a fast way for SSP measurement that can be performed during ship navigation, and the time cost is related to the measurement depth. For HYLMT-2000 used in this experiment, it takes only about 20 minutes to measure an SSP with maximum depth of 2000 meters. These 13 SSPs were collected as reference SSPs. The full--depth of SSP measured by CTD at the area center for testing was set as an SSP prediction task. The location of anchor nodes and sampled SSPs are shown in Fig.\ref{fig_2b}.

\subsection{Experimental Settings}
\subsubsection{Data source}  
\begin{figure}[!t]
	\centering
	\includegraphics[width=3in]{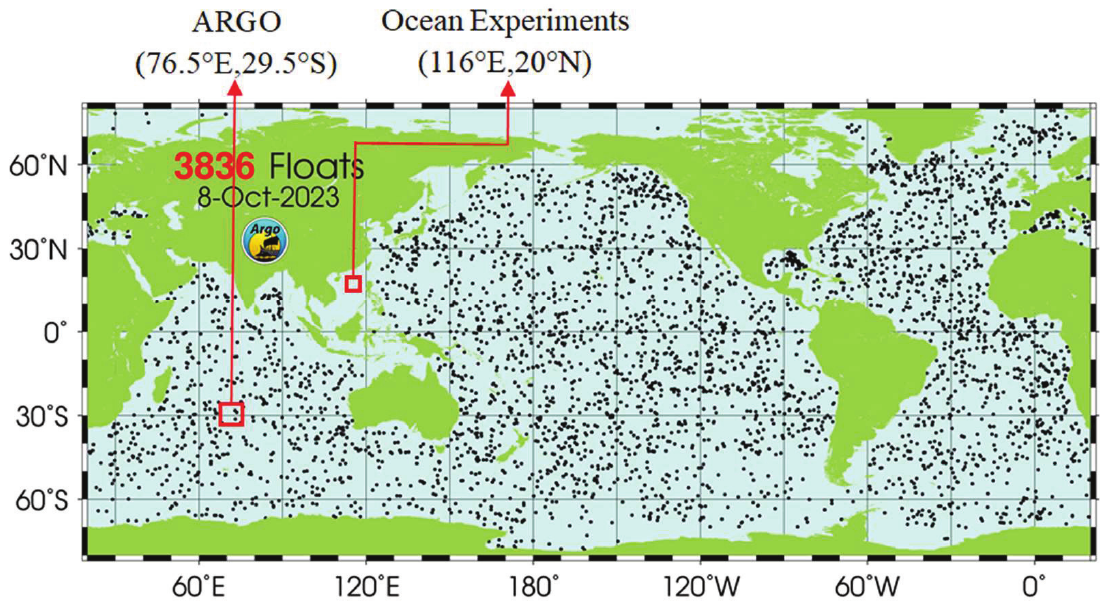}
	\caption{Spatial positions of ARGO and Ocean Experiments SSP samples.}
	\label{fig_3}
\end{figure}

To evaluate the performance of proposed H--LSTM, SSP data from the ARGO website \cite{Zhang2021Argo} and ocean experiments are used for model testing. The sampling locations for ARGO SSP data and ocean experiment SSP data are shown in Fig.~\ref{fig_3}, and Table~\ref{tab_1} provides the data information used in this experiment.

\subsubsection{Platform and parameters}

The simulations were all implemented in MATLAB R2021a. There are 128 hidden layer neurons. The learning rate is 0.01, and the number of epochs is 300.

\subsection{Accuracy Performance of H-LSTM}
In this section, we first analyze the time series SSP prediction performance of H--LSTM, then compare its accuracy performance with the state--of--the--art methods: the mean value prediction method, polynomial fitting method \cite{Liu2019Polynomial}, and BP neural network \cite{Yu2020RBF}.

For the ARGO SSP data, taking the prediction start time as October 2021 as an example, 48 historical layered sound speed distribution data from October 2017 to September 2021 were used as learning samples for H--LSTM. The comparison between the original and the predicted 58 layers' sound speed data for February and October in 2021 is shown in Fig. \ref{fig_4} (a-b).

Due to the fact that the historical hierarchical sound speed distribution time standardized dataset is not full-ocean depth data, but layered data, linear interpolation method is used to interpolate the predicted layered SSP and verified layered SSP within the full-ocean depth range. The comparison between the predicted full-ocean depth SSP in February and October and the actual full-ocean depth SSP is shown in Fig. \ref{fig_4}(c-d). It shows that the H-LSTM network model can make accurate predictions of the future full-ocean depth SSP in month.

In order to more intuitively evaluate the performance of the H-LSTM model, we present the prediction errors for different depth layers in Table \ref{tab_2}. It can be seen that the prediction error of each depth layer is less than 1m/s, and among the 36 layers (24 layers are shown), only 3 layers have RMSE>0.3m/s.

\begin{table}[!t]
	\caption{PREDICTION ERRORS OF DIFFERENT DEOTH LAYERS}\label{tab_2}   
	\centering
	\begin{tabular}{|c|c|c|c|c|c|}
		\hline
		\makecell[c]{Depth\\Layers}&\makecell[c]{RMSE\\(m/s)}&\makecell[c]{Depth\\Layers}&\makecell[c]{RMSE\\(m/s)}&\makecell[c]{Depth\\Layers}&\makecell[c]{RMSE\\(m/s)} \\  
		\hline
		1&0.0545&13&0.1730&25&0.0683\\
		\hline		
		2&0.8341&14&0.2027&26&0.0864\\
		\hline		
		3&0.5097&15&0.1073&27&0.0663\\
		\hline		
		4&0.0567&16&0.1504&28&0.0663\\
		\hline
		6&0.4674&18&0.0046&30&0.1065\\
		\hline		
		8&0.0592&20&0.0104&32&0.0420\\
		\hline		
		10&0.0698&22&0.0904&34&0.0397\\
		\hline		
		12&0.2345&24&0.0426&36&0.0630\\
		\hline
	\end{tabular}
\end{table}

Table \ref{tab_3} shows the RMSE between the predicted full-ocean depth SSP and the actual full-ocean depth SSP using the H-LSTM model with ARGO data and ocean experiments data. It can be intuitively seen from the table that the H-LSTM neural network model performs well, with the RMSE less than 0.5m/s.

\begin{table}[!t]
	\caption{ERRORS BETWEEN PREDICTED AND ACTUAL FULL OCEAN DEPTH SSP}\label{tab_3}   
	\centering
	\begin{tabular}{|c|c|c|c|}
		\hline
		\multicolumn{2}{|c|}{\makecell[c]{\textbf{Argo}\\\textbf{Dataset}}}&\multicolumn{2}{c|}{\makecell[c]{\textbf{Ocean Experiments}\\ \textbf{Dataset}}}\\
		\hline
		\makecell[c]{Predicted\\Area}&\makecell[c]{Indian Ocean\\(76.5$^{\circ}$E,29.5$^{\circ}$S)}&\makecell[c]{Predicted\\Area}&\makecell[c]{South China Sea\\(116$^{\circ}$E,20$^{\circ}$N)}\\
		\hline		
		\makecell[c]{Full\\Depth}&0-1975m&\makecell[c]{Full\\Depth}&0-3500m\\
		\hline		
		\makecell[c]{Predicted\\time}&\makecell[c]{RMSE\\(m/s)}&\makecell[c]{Predicted\\time}&\makecell[c]{RMSE\\(m/s)}\\
		\hline		
		2021.02&\textbf{0.2637}&\multirow{4}*{\makecell[c]{2023.03.27\\24:00}}&\multirow{4}*{\textbf{0.1530}}\\
		\cline{1-2}	
		2021.10&0.3063&\multicolumn{1}{c|}{}&\multicolumn{1}{c|}{}\\
		\cline{1-2}	
		2021.12&0.4494&\multicolumn{1}{c|}{}&\multicolumn{1}{c|}{}\\
		\cline{1-2}			
		\makecell[c]{Average\\RMSE}&0.3398&\multicolumn{1}{c|}{}&\multicolumn{1}{c|}{}\\
		\hline			
	\end{tabular}
\end{table}

\begin{figure}[!htbp]   
	\centering
	\subfloat[]{\includegraphics[width=1.5in]{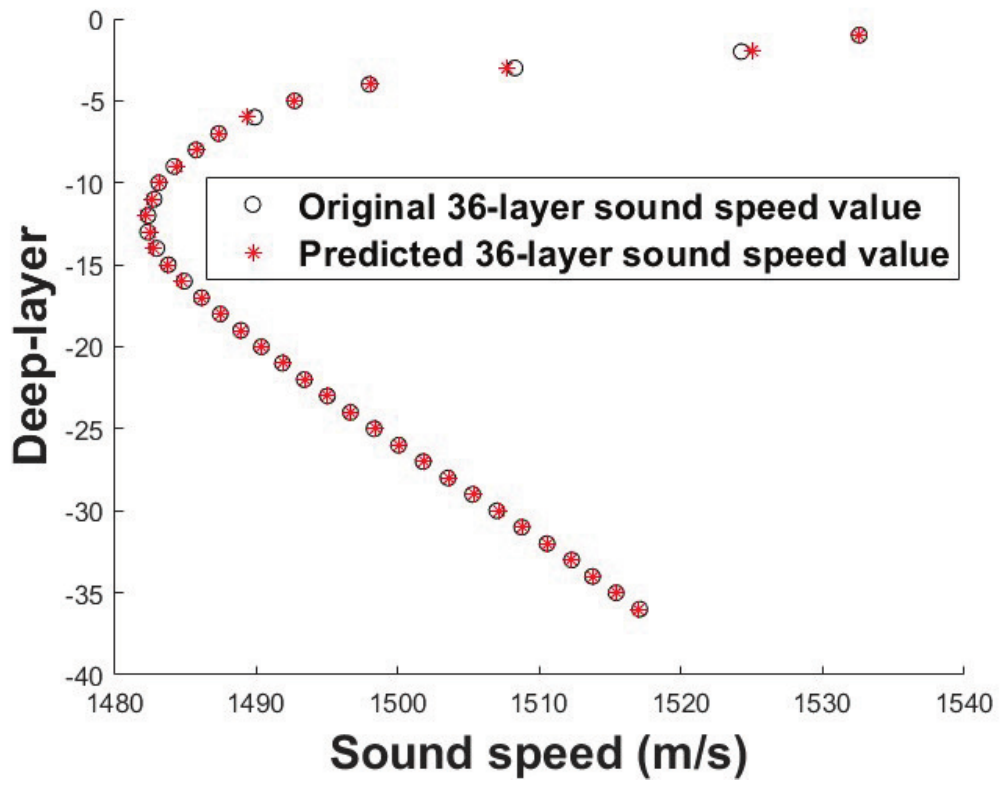}%
		\label{fig_5_1}}
	\hfil
	\subfloat[]{\includegraphics[width=1.5in]{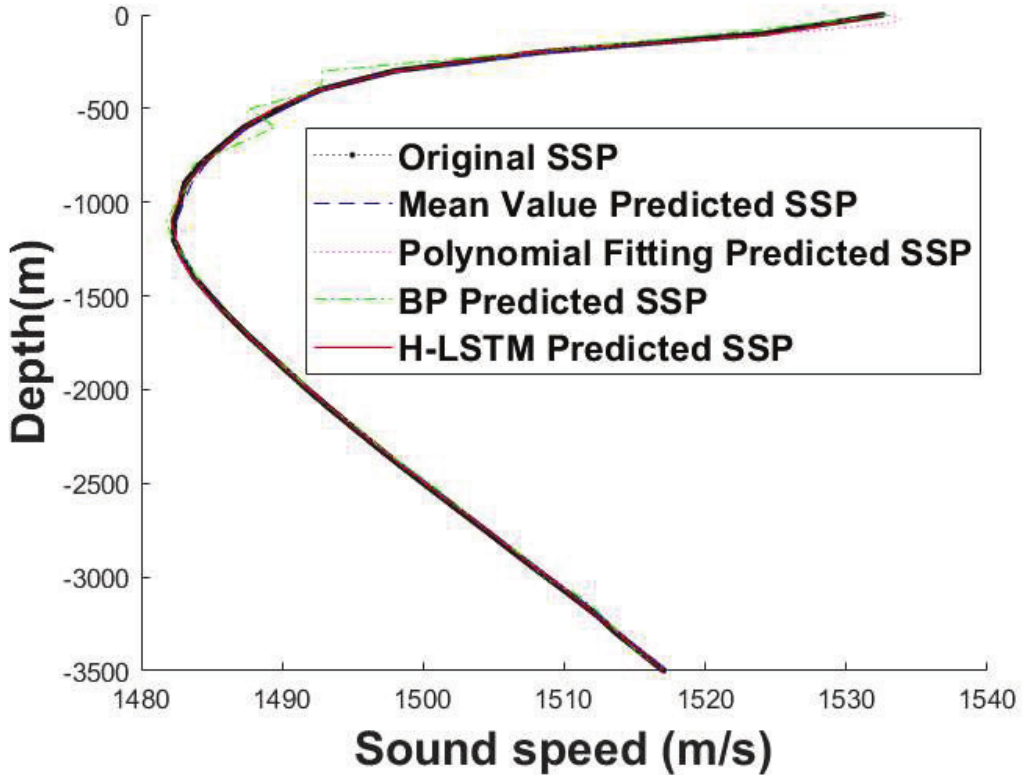}%
		\label{fig_5_2}}	
	\caption{Comparison between predicted hierarchical SSP and actual hierarchical SSP (Ocean Experiments Dataset). (a)Comparison of layered data. (b) Comparison with different state--of--the--art methods. }
	\label{fig_5}
\end{figure}

For the hierarchical ocean experiments data, the model is trained using 13 SSP data before the prediction start time. The comparison between the predicted average hierarchical SSP in future two hours and the actual hierarchical SSP is shown in Fig.~\ref{fig_5_1}. The comparison between the predicted full-ocean depth SSP using four methods and the actual full-ocean depth SSP is shown in Fig. \ref{fig_5_2}. The polynomial fitting method performs polynomial fitting on the historical SSP data of two consecutive years before February 2021 to obtain the predicted future SSP. The BP neural network uses the same training dataset as H-LSTM as the model's training set to predict future SSP.

From Fig. \ref{fig_5_2}, it can be seen that for the mean value prediction method and polynomial fitting prediction method, it can roughly fit the main features of future SSP smoothly, but there are some numerical deviations overall. For the BP neural network prediction method, its predicted future SSP is not smooth enough, fluctuating repeatedly around the actual SSP, losing the main features of the actual SSP, and this result is not suitable as a prediction result for future SSP. For the H-LSTM neural network prediction method we proposed, compared with the actual SSP, the predicted full-ocean depth SSP does not lose the main features of the actual SSP, nor does it have an overall numerical deviation problem. In both shallow and deep-sea areas, the prediction results are relatively good, with only minor numerical deviations in some details.

We provide a comparison of the RMSE results of four methods in training and predicting future full-ocean depth SSPs using two different datasets in Table \ref{tab_4}. From the data in the table, it can be seen that the H-LSTM model performs best in predicting using the ARGO dataset and the ocean experiment dataset, with RMSE of 0.2637m/s and 0.1530m/s, respectively. Through comparative experiments, we have fully verified the excellent performance of H-LSTM neural network model in predicting future full-ocean depth SSP processes.

\begin{table}[!htbp]
	\caption{RMSE RESULTS OF FOUR METHODS}\label{tab_4}   
	\centering
	\begin{tabular}{|c|c|c|c|c|c|c|c|c|c|}
		\hline
		\multicolumn{10}{|c|}{\textbf{Argo Dataset}}\\
		\hline
		\multicolumn{5}{|c}{Predicted Area}&\multicolumn{5}{|c|}{Indian Ocean (76.5$^{\circ}$E,29.5$^{\circ}$S)}\\
		\hline	
		\multicolumn{5}{|c}{Predicted time}&\multicolumn{5}{|c|}{February 2021}\\
		\hline			
		\multicolumn{5}{|c}{Full Depth}&\multicolumn{5}{c|}{0-1975m}\\
		\hline
		\multicolumn{2}{|c}{Method}&\multicolumn{2}{|c}{\makecell[c]{Mean value\\ predicted}}&\multicolumn{2}{|c}{\makecell[c]{Polynomial\\ Fitting}}&\multicolumn{2}{|c}{BP}&\multicolumn{2}{|c|}{H-LSTM}\\	
		\hline
		\multicolumn{2}{|c}{RMSE(m/s)}&\multicolumn{2}{|c}{1.0082}&\multicolumn{2}{|c}{0.9134}&\multicolumn{2}{|c}{1.4854}&\multicolumn{2}{|c|}{\textbf{0.2637}}\\	
		\hline	
		\hline		
		\multicolumn{10}{|c|}{\textbf{Ocean Experiments Dataset}}\\
		\hline
		\multicolumn{5}{|c}{Predicted Area}&\multicolumn{5}{|c|}{South China Sea (116$^{\circ}$E,20$^{\circ}$N)}\\
		\hline	
		\multicolumn{5}{|c}{Predicted time}&\multicolumn{5}{|c|}{March 27, 2023 24:00}\\
		\hline		
		\multicolumn{5}{|c}{Full Depth}&\multicolumn{5}{c|}{0-3500m}\\
		\hline
		\multicolumn{2}{|c}{Method}&\multicolumn{2}{|c}{\makecell[c]{Mean value\\ predicted}}&\multicolumn{2}{|c}{\makecell[c]{Polynomial\\ Fitting}}&\multicolumn{2}{|c}{BP}&\multicolumn{2}{|c|}{H-LSTM}\\	
		\hline
		\multicolumn{2}{|c}{RMSE(m/s)}&\multicolumn{2}{|c}{0.2835}&\multicolumn{2}{|c}{0.5480}&\multicolumn{2}{|c}{0.9573}&\multicolumn{2}{|c|}{\textbf{0.1530}}\\	
		\hline			
	\end{tabular}
\end{table}

\subsection{H-LSTM’s performance in predicting cyclical changes in SSPs}
\begin{figure}[!t]
	\centering
	\includegraphics[width=2.5in]{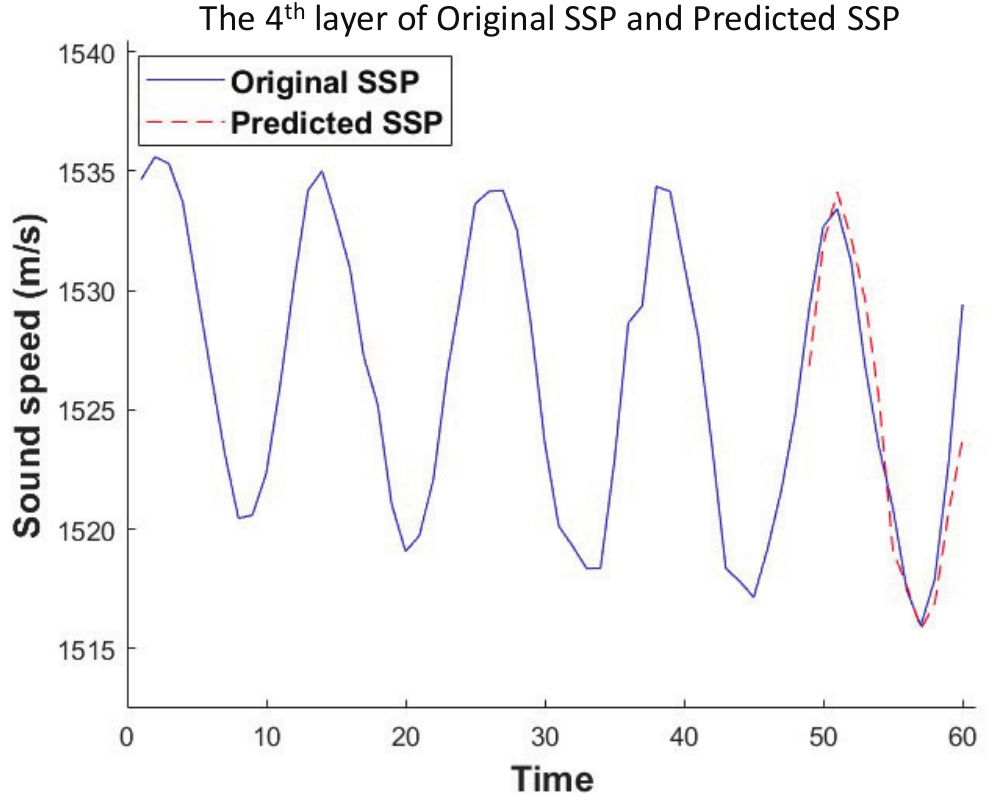}
	\caption{Comparison between predicted full-ocean depth SSP using four methods and original full-ocean depth SSP(ARGO Dataset).}
	\label{fig_6}
\end{figure}

To test the performance of the H-LSTM model in predicting periodic changes in SSP, we used historical SSPs data from 2017 to 2020 as learning samples and conducted a 12 steps prediction on future SSP data, namely predicting SSPs for all months of the next year. To ensure temporal rigor, as a single SSP data is consistent in time. Therefore, the fourth layer of SSP data (corresponding to a depth of 20 meters) is taken from the layered data to test whether this method can accurately capture the periodic changes of sound speed distribution. The schematic diagrams for comparing the periodic change trends of the predicted data in the third depth layers with the original SSP data periodic change trends are shown in Fig. \ref{fig_6}.

The 60 solid blue line data in the figure represent the actual SSP data for the 60 months from 2017 to 2021 at the corresponding depth layer, with the first 48 being training data and the last 12 being validation data. The 12 red dashed line data represent the predicted SSP data for the next 12 months at the corresponding depth layer, which is compared with the validation data. It can be intuitively seen that our proposed H-LSTM method can accurately capture the periodic changes of SSP over time.

\section{Conclusion}   
To fast estimate the future distribution of SSP, we propose an H-LSTM method to dynamically predict future full-ocean depth SSP. In order to verify the feasibility of the model, we conducted experimental simulations on two different datasets for training, predicting the average full-ocean depth SSP for the next month and the next two hours, and setting different methods as control experiments. The experimental results shows that the proposed H-LSTM model can not only make accurate predictions of future full-ocean depth SSP, but also accurately capture the periodic changes of SSP over time.

\section*{Acknowledgments}
This work was supported by Natural Science Foundation of Shandong Province (ZR2023QF128), China Postdoctoral Science Foundation (2022M722990), Qingdao Postdoctoral Science Foundation (QDBSH20220202061), National Natural Science Foundation of China (62271459), National Defense Science and Technology Innovation Special Zone Project: Marine Science and Technology Collaborative Innovation Center (22-05-CXZX-04-01-02), Fundamental Research Funds for the Central Universities, Ocean University of China (202313036).

\bibliographystyle{IEEEtran}
\bibliography{IEEE_Letters_draft}


\end{document}